\documentclass[11pt,a4paper]{article}

\usepackage{jheppub,bm}

\title{Unruh effect for neutrinos interacting with accelerated matter}

\author[a,b]{Maxim Dvornikov}
\affiliation[a]{Physics Faculty, National Research Tomsk State University,
36 Lenin Ave., 634050 Tomsk, Russia}
\affiliation[b]{Pushkov Institute of Terrestrial Magnetism, Ionosphere
and Radiowave Propagation (IZMIRAN),
142190 Troitsk, Moscow, Russia}
\emailAdd{maxdvo@izmiran.ru}

\abstract{
We study the evolution of neutrinos in a background matter moving
with a linear acceleration. The Dirac equation for a massive neutrino
electroweakly interacting with background fermions is obtained in
a comoving frame where matter is at rest. We solve this Dirac equation
for ultrarelativistic neutrinos. The neutrino quantum states in matter
moving with a linear acceleration are obtained. We demonstrate that the neutrino
electroweak interaction with an accelerated matter leads to the vacuum
instability which results in the neutrino-antineutrino pairs creation.
We rederive the temperature of the Unruh radiation and find the correction
to the Unruh effect due to the specific neutrino interaction with
background fermions. As a possible
application of the obtained results we discuss the neutrino pairs creation
in a core collapsing supernova. The astrophysical upper limit on the neutrino masses
is obtained.
}


\arxivnumber{1507.01174}

\begin{document}

\maketitle

\section{Introduction}

The standard model neutrino interaction with other fermions is known
to be extremely weak. Nevertheless, in some cases, even this weak
interaction is crucial in the evolution of a neutrino system. One
can recall the resonant amplification of neutrino flavor oscillations
in background matter -- the Mikheyev-Smirnov-Wolfenstein (MSW) effect~\cite{MikSmi85,Wol78}
-- which is the most plausible explanation of the solar neutrinos deficit~\cite{HaxRobSer13}.
Typically background matter, which a neutrino interacts with, is considered
to either be at rest or move with a constant velocity. In both cases
matter effects can influence the neutrino oscillations picture.

Recently neutrino flavor oscillations in matter moving with an acceleration
was discussed in refs.~\cite{FraNau13,StuTok14,Dvo14}.
The modification of the neutrino refraction index owing to a nonzero
matter acceleration was accounted for in refs.~\cite{FraNau13,StuTok14}
without consideration of possible noninertial effects. The neutrino motion
in an accelerated frame accounting for noninertial effects was described
in ref.~\cite{Dvo14}. The influence
of the Unruh radiation on neutrino flavor oscillations was discussed
in ref.~\cite{AhlLabTor15}.

Linearly accelerated particles reveal another interesting effect consisting in the emission of a thermal radiation, which was
first predicted by Unruh~\cite{Unr76}. The temperature of this radiation
appears to be proportional to the particle acceleration. This effect
is still widely discussed in the literature (for a recent review see,
e.g., ref.~\cite{CriHigMat08}). In ref.~\cite{BieMer06}, it was
shown that the Unruh effect strongly depends on the way to observe a thermal radiation. Nevertheless there are some suggestions how
to detect the Unruh radiation experimentally~\cite{MarFueMan11}.

In the present work we continue to study the influence of noninertial effects on electroweakly
interacting particles. In refs.~\cite{Dvo14,Dvo15} we considered
the electroweak interaction of neutrinos, electrons, and quarks with
a rotating background matter in a corotating frame. In particular,
the new galvano-rotational effect was predicted in ref.~\cite{Dvo15}.
In this work we study the neutrino interaction with a linearly accelerated
matter. Note that this kind of the matter acceleration can be implemented in various astrophysical
media such as a supernova (SN) at the bounce stage~\cite{ThoBurPin03}
and jets from active galactic nuclei~\cite{Spu15}. One can also
expect intense neutrino beams in these environments.

This work is organized as follows. In section~\ref{sec:NUMATTFLAT}
we recall the standard model neutrino interaction with background
fermions which can move with a constant velocity. In section~\ref{sec:DIRACEQ}
we derive and solve the Dirac equation, accounting for noninertial
effects, for ultrarelativistic neutrinos interacting with a linearly
accelerated matter. In section~\ref{sec:CLASSQUANTST} we study the
neutrino quantum states in a linearly accelerated matter. The neutrino-antineutrino
($\nu\bar{\nu}$) pairs creation is described in section~\ref{sec:PAIRSCR}.
In paticular, we reproduce the Unruh effect for neutrinos in electroweakly interacting
matter moving with a linear acceleration and consider the correction
to this effect owing to the specific neutrino interaction. In section~\ref{sec:APPL}
we propose the astrophysical application of our results, consisting in the $\nu\bar{\nu}$ pairs creation
in a SN explosion at the bounce stage. Finally, in section~\ref{sec:CONCL}
we discuss our results. Some useful mathematical formulas are provided in appendix~\ref{sec:WMPROP}.

\section{Neutrino interaction with matter in an inertial frame\label{sec:NUMATTFLAT}}

In this section we shall briefly remind how active neutrinos interact
with background matter in the flat space-time. We describe the neutrino
electroweak interaction in the Fermi approximation.

The electroweak interaction of the flavor neutrino eigenstates $\nu_{\alpha}$,
$\alpha=e,\mu,\tau$, with a background matter consisting of electrons
$e$, protons $p$, and neutrons $n$, can be described in the mean
field approximation by the Langrangian
\begin{equation}\label{eq:LagrFE}
  \mathcal{L}_{\mathrm{int}} =
  -\sum_{\alpha=e,\mu,\tau}
  \bar{\nu}_{\alpha}\gamma_{\mu}^{\mathrm{L}}\nu_{\alpha}
  \cdot
  J_{\nu_{\alpha}}^{\mu},
\end{equation}
where $\gamma_{\mu}^{\mathrm{L}}=\gamma_{\mu}\left(1-\gamma^{5}\right)/2$,
$\gamma^{\mu}=\left(\gamma^{0},\bm{\gamma}\right)$ are the Dirac
matrices, and $\gamma^{5}=\mathrm{i}\gamma^{0}\gamma^{1}\gamma^{2}\gamma^{3}$.
The effective current in eq.~(\ref{eq:LagrFE}) has the form~\cite{DvoStu02},
\begin{equation}\label{eq:effcurr}
  J_{\nu_{\alpha}}^{\mu}=\sqrt{2}G_{\mathrm{F}}\sum_{f=e,p,n}\left(q_{f}^{(1)}j_{f}^{\mu}+q_{f}^{(2)}\lambda_{f}^{\mu}\right),
\end{equation}
where $G_{\mathrm{F}}=1.17\times10^{-5}\thinspace\text{GeV}^{-2}$
is the Fermi constant, $j_{f}^{\mu}$ is the hydrodynamic current
of background fermions, $\lambda_{f}^{\mu}$ is the four vector of
the mean polarization. The hydrodynamic current $j_{f}^{\mu}=n_{f}u_{f}^{\mu}$
depends on the invariant number density $n_{f}$, i.e. the density
in the rest frame of background fermions, and the macroscopic mean
four velocity $u_{f}^{\mu}$ of these fermions. The explicit form
of $\lambda_{f}^{\mu}$ in terms of the invariant density, four velocity,
and the invariant polarization $\bm{\zeta}_{f}$ can be found, e.g.,
in ref.~\cite{Dvo13}. The coefficients $q_{f}^{(1,2)}$ in eq.~(\ref{eq:effcurr})
for $\nu_e$ have the form~\cite{DvoStu02}
\begin{equation}\label{eq:q1q2nue}
  q_{1}^{(f)} =
  I_{\mathrm{L}3}^{(f)}-2Q_{f}\xi+\delta_{ef},
  \quad
  q_{2}^{(f)}=-I_{\mathrm{L}3}^{(f)}-\delta_{ef},
\end{equation}
where $I_{\mathrm{L}3}^{(f)}$ is the third component of the weak
isospin of type $f$ fermions, $Q_{f}$ is the value of their electric
charge, $\xi=\sin^{2}\theta_{\mathrm{W}} \approx 0.23$ is the Weinberg parameter, $\delta_{ef}=1$
for electrons and vanishes for protons and neutrons. To get $q_{f}^{(1,2)}$
for $\nu_{\mu,\tau}$ interactions with the same background fermions,
we should set $\delta_{ef}=0$ in eq.~(\ref{eq:q1q2nue}).

The Lagrangian in eq.~(\ref{eq:LagrFE}) can be obtained by the averaging
of the fermionic vector and axial-vector currents in the rest frame
of background fermions. Therefore, fermions, contributing to eq.~(\ref{eq:effcurr}),
are supposed to have constant and homogeneous mean velocities. Otherwise
(e.g, for matter moving with an acceleration; cf. section~\ref{sec:DIRACEQ}
below), eq.~(\ref{eq:effcurr}) is not valid. Indeed to obtain eq.~(\ref{eq:effcurr})
one should make a Lorentz boost to the rest frame of background fermions
to make the averaging and then make a boost back to the laboratory
frame. Such coordinate transformations are undefined if the Lorentz
symmetry is broken.

In our work we shall be mainly interested in the neutrino evolution
in the dense electrically neutral nuclear matter of a protoneutron
star (PNS). Let us assume that this matter is at rest and unpolarized. In this case, on the basis of eqs.~(\ref{eq:effcurr}) and~(\ref{eq:q1q2nue})
we get the effective potential $V_{\nu_{\alpha}}=J_{\nu_{\alpha}}^{0}$
as
\begin{equation}\label{eq:Vnue}
  V_{\nu_{\alpha}}\approx-\frac{G_{\mathrm{F}}}{\sqrt{2}}n_{n},
\end{equation}
where $n_{n}$ is the neutron density. Note that the effective potentials
for different neutrino types are approximately equal since the densities
of charged particles inside PNS are much lower than
that of neutrons.

In the wake of the results of the recent experiments (see, e.g., ref.~\cite{An15}),
it is commonly believed that massive neutrino eigenstates, i.e. those
which have definite masses, are the mixture of flavor neutrinos $\nu_{\alpha}$
defined above. It should be noted that the matter interaction of mass
eigenstates is nondiagonal in the particle species. In the present
work we shall not distinguish between mass and flavor eigenstates.
For instance, this approximation is valid if we study the evolution of the $\nu_{e}-\nu_{\tau}$
system, where the mixing angle $\theta_{13}$ is known to be the smallest. Alternatively we may consider the neutrino system with the parameters far away from any MSW resonance.

According to many theoretical models for the neutrino mass generation~\cite{BouMorVal14},
the neutrino mass eigenstates are likely to be Majorana particles.
Nevertheless, in the present work, we shall assume that the neutrino
mass eigenstates $\psi_{i}$, $i=1,2,\dotsc$, are the Dirac particles.
Note that, despite the great experimental efforts~\cite{Alb14},
the issue whether neutrinos are Dirac or Majorana particles still
remains open.

\section{Dirac equation for a neutrino interacting with linearly accelerated
matter\label{sec:DIRACEQ}}

In this section we shall describe the motion of a neutrino in a linearly
accelerated matter. Our analysis will be based on the exact solution
of the Dirac equation accounting for noninertial effects.

The neutrino interaction with background matter described in section~\ref{sec:NUMATTFLAT}
implied the uniform matter motion with a constant velocity, i.e. $J_{\nu_{\alpha}}^{\mu}$
in eq.~(\ref{eq:effcurr}) depends on neither coordinates nor time.
The opposite situation would mean an accelerated matter. As was mentioned
in section~\ref{sec:NUMATTFLAT}, for accelerated matter we cannot
assume a nonzero $\mathbf{J}_{\nu_{\alpha}}(\mathbf{r},t)$ in eq.~(\ref{eq:LagrFE})
and in the corresponding Dirac equation, written in the flat space-time,
since the Lorentz symmetry is broken.

Nevertheless, even for accelerated matter, we can always find a reference
frame where matter is at rest. Assuming that matter is unpolarized,
we get that only $J_{\nu_{\alpha}}^{0}\neq0$
exists in this reference frame. Thus, in this frame, we can still
use the effective potential defined in eq.~(\ref{eq:Vnue}) since
it depends on the invariant density of background fermions. However,
in this case the Dirac equation should be modified to account for
noninertial effects. This approach was developed in refs.~\cite{Dvo14,Dvo15},
where we studied the propagation of electroweakly interacting particles
(including neutrinos) in a rotating matter.

It is known that the description of the particle motion by an accelerated
observer is equivalent to the interaction of this particle with an effective
gravitational field. Hence, in our treatment of the neutrino evolution,
we should rewrite the Dirac equation in the curved space-time corresponding
to a linearly accelerated frame, i.e in the Rindler space-time. If
we suppose that the matter is accelerated along the $z$-axis, the
interval in the Rindler space-time has the form~\cite{MisThoWhe73},
\begin{equation}\label{eq:mertrot}
  \mathrm{d}s^{2} =
  g_{\mu\nu}\mathrm{d}x^{\mu}\mathrm{d}x^{\nu} =
  a^{2}z^{2}\mathrm{d}t^{2} -
  \mathrm{d}x^{2}-\mathrm{d}y^{2}-\mathrm{d}z^{2},
\end{equation}
where $g_{\mu\nu}$ is the metric tensor of the effective gravitational
field and $a$ is the proper acceleration of matter. Here we use
the Cartesian coordinates $x^{\mu}=(t,x,y,z)$. The interval in eq.~(\ref{eq:mertrot})
can be transformed to the Minkowskian one $\mathrm{d}s^{2}=\mathrm{d}\mathcal{T}^{2}-\mathrm{d}\mathcal{X}^{2}-\mathrm{d}\mathcal{Y}^{2}-\mathrm{d}\mathcal{Z}^{2}$
by changing the coordinates
\begin{alignat}{5}
  \mathcal{T} = & z\sinh at, &
  \qquad
  \mathcal{Z} = & z\cosh at, &
  \qquad
  \mathcal{X} = & x,
  \qquad
  \mathcal{Y} = & y.
  \notag
  \\
  t = & \frac{1}{a}\text{arctanh}\frac{\mathcal{T}}{\mathcal{Z}}, &
  \qquad
  z = & \sqrt{\mathcal{Z}^{2}-\mathcal{T}^{2}}, &
  \qquad
  x = & \mathcal{X},
  \qquad
  y = & \mathcal{Y}.
\end{alignat}
It should be noted that the metric in eq.~(\ref{eq:mertrot}) does
not span over the all space-time. It covers only the sector $\mathcal{Z}>|\mathcal{T}|$
in the $(\mathcal{Z},\mathcal{T})$ hyperplane.

The Dirac equation for a neutrino interacting with background
matter in a curved space-time can be obtained, using the results of refs.~\cite{Dvo14,Dvo15,GriMamMos80}
and accounting for eq.~(\ref{eq:LagrFE}), in the form,
\begin{equation}\label{eq:Depsicurv}
  \left[
    \mathrm{i}\gamma^{\mu}(x)\nabla_{\mu}-m
  \right]
  \psi =
  \frac{1}{2}J^{\mu}\gamma_{\mu}(x)
  \left[
    1-\gamma^{5}(x)
  \right]
  \psi,
\end{equation}
where $\psi$ is the neutrino bispinor, $m$ is the neutrino mass,
$\gamma^{\mu}(x)$ are the coordinate dependent Dirac matrices, $\nabla_{\mu}=\partial_{\mu}+\Gamma_{\mu}$
is the covariant derivative, $\Gamma_{\mu}$ is the spin connection,
$\gamma^{5}(x)=-(\mathrm{i}/4!)E^{\mu\nu\alpha\beta}\gamma_{\mu}(x)\gamma_{\nu}(x)\gamma_{\alpha}(x)\gamma_{\beta}(x)$,
$E^{\mu\nu\alpha\beta}=\varepsilon^{\mu\nu\alpha\beta}/\sqrt{-g}$
is the covariant antisymmetric tensor in curved space-time, $g=\det(g_{\mu\nu})$,
and $J^{\mu}$ is the effective external current of background fermions;
cf. eq.~(\ref{eq:effcurr}). Note that, since we choose a comoving
frame, only the zeroth component of $J^{\mu}$ is nonvanishing: $J^{0}=V\neq0$,
where $V$ is given in eq.~(\ref{eq:Vnue}). In eq.~(\ref{eq:Depsicurv})
we omit the index $\nu_{\alpha}$ for brevity: $V \equiv V_{\nu_{\alpha}}$,
$m\equiv m_{\nu_{\alpha}}$ etc.

One can check that the metric tensor in eq.~(\ref{eq:mertrot}) can
be diagonalized using the following vierbein vectors:
\begin{equation}\label{eq:vierbein}
  e_{0}^{\ \mu} =
  \left(
    \frac{1}{az},0,0,0
  \right),
  \quad
  e_{1}^{\ \mu} = (0,1,0,0),
  \quad
  e_{2}^{\ \mu} = (0,0,1,0),
  \quad
  e_{3}^{\ \mu} = (0,0,0,1).
\end{equation}
Indeed, the direct calculation using eq.(\ref{eq:vierbein}) shows
that $\eta_{ab}=e_{a}^{\ \mu}e_{b}^{\ \nu}g_{\mu\nu}$, where $\eta_{ab}=\text{diag}(1,-1,-1,-1)$
is the metric in a locally Minkowskian frame.

To obtain the spin connection one should introduce the constant Dirac
matrices in a locally Minkowskian frame as $\gamma^{\bar{a}}=e_{\ \mu}^{a}\gamma^{\mu}(x)$.
As shown in ref.~\cite{Dvo14}, $\gamma^{5}(x) = \mathrm{i} \gamma^{\bar{0}} \gamma^{\bar{1}} \gamma^{\bar{2}} \gamma^{\bar{3}} = \gamma^{\bar{5}}$
does not depend on coordinates. The spin connection in the Dirac eq.~(\ref{eq:Depsicurv})
has the form~\cite{GriMamMos80},
\begin{equation}\label{eq:spincon}
  \Gamma_{\mu}=-\frac{\mathrm{i}}{4}\sigma^{ab}\omega_{ab\mu},
  \quad
  \omega_{ab\mu}=e_{a}^{\ \nu}e_{b\nu;\mu},
\end{equation}
where $\sigma_{ab}=(\mathrm{i}/2)[\gamma_{\bar{a}},\gamma_{\bar{b}}]_{-}$ are
the generators of the Lorentz transformations in a locally Minkowskian
frame and the semicolon stays for the covariant derivative. The explicit
calculation on the basis of eq.~(\ref{eq:spincon}) shows that the
nonzero components of the connection one-form $\omega_{ab}=\omega_{ab\mu}\mathrm{d}x^{\mu}$
are
\begin{equation}\label{eq:spinconcomp}
  \omega_{01\mu}=-\omega_{10\mu}=(a,0,0,0).
\end{equation}
Using eqs.~(\ref{eq:spincon}) and~(\ref{eq:spinconcomp}) we get
that $\mathrm{i}\gamma^{\mu}(x)\Gamma_{\mu}=\mathrm{i}\gamma^{\bar{3}}/2z$.

Using the definition of $\gamma^{\bar{a}}$, the Dirac eq.~(\ref{eq:Depsicurv})
takes the form,
\begin{equation}\label{eq:Direqgen}
  \left[
    \mathrm{i}\gamma^{\bar{0}} \frac{\partial_{0}}{az} +
    \mathrm{i}\gamma^{\bar{1}}\partial_{x} +
    \mathrm{i}\gamma^{\bar{2}}\partial_{y} +
    \mathrm{i}\gamma^{\bar{3}}
    \left(
      \partial_{z}+\frac{1}{2z}
    \right) -
    m
  \right]
  \psi =
  \frac{1}{2} az\gamma^{\bar{0}}V
  \left(
    1-\gamma^{\bar{5}}
  \right)
  \psi.
\end{equation}
Analogous Dirac equation in the Rindler wedge with $V=0$ was studied
in ref.~\cite{SofMulGre80}. Since eq.~(\ref{eq:Direqgen}) does
not explicitly depend on $t$, $y$, and $z$, we shall look for its
solution in the form,
\begin{equation}\label{eq:psipsir}
  \psi =
  \exp
  \left(
    -\mathrm{i}Et+\mathrm{i}p_{x}x+\mathrm{i}p_{y}y
  \right)
  \psi_{z},
\end{equation}
where $\psi_{z}=\psi_{z}(z)$ is the wave function depending on $z$.

It is convenient to rewrite eq.~(\ref{eq:Direqgen}) as
\begin{equation}\label{eq:psir}
  \left[
    \gamma^{\bar{a}}Q_{a}-m+U
  \right]
  \psi_{z} = 0,
\end{equation}
where $Q^{a}=q^{a}-q_{\mathrm{eff}}A_{\mathrm{eff}}^{a}$, $q_{\mathrm{eff}}$
is the effective electric charge, $q^{a}=\left(0,p_{x},p_{y},-\mathrm{i}\partial_{z}\right)$,
\begin{equation}
  A_{\mathrm{eff}}^{a} =
  \frac{1}{q_{\mathrm{eff}}}
  \left(
    \frac{azV}{2}-\frac{E}{az},0,0,\frac{\mathrm{i}}{2z}
  \right),
\end{equation}
is the potential of the effective electromagnetic field, and $U = azV\gamma^{\bar{0}}\gamma^{\bar{5}}/2$.

Let us look for the solution of eq.~(\ref{eq:psir}) in the form,
$\psi_{z}=\left[\gamma^{\bar{a}}Q_{a}+m-U\right]\Phi$, where $\Phi$
is the new spinor. The equation for $\Phi$ reads
\begin{multline}\label{eq:Phieq}
  \bigg[
    \left(
      \partial_{z}+\frac{1}{2z}
    \right)^{2} +
    \left(
      \frac{E}{az}-\frac{azV}{2}
    \right)^{2} -
    p_{\perp}^{2}+\frac{a^{2}z{}^{2}V^{2}}{4}-m^{2}
    \\
    +
    \left(
      \frac{E}{az{}^{2}}+\frac{aV}{2}
    \right)
    \mathrm{i}\alpha_{3}-\frac{aV}{2}
    \left[
      2z
      \left(
        \frac{E}{az}-\frac{azV}{2}
      \right) -
      \mathrm{i}\alpha_{3}
    \right]
    \gamma^{\bar{5}} +
    mazV\gamma^{\bar{0}}\gamma^{\bar{5}}
  \bigg]
  \Phi = 0,
\end{multline}
where $p_{\perp}^{2}=p_{x}^{2}+p_{y}^{2}$, $\alpha_{3}=\gamma^{\bar{5}}\Sigma_{3}$,
and $\Sigma_{3}=\gamma^{\bar{0}}\gamma^{\bar{3}}\gamma^{\bar{5}}$.

The solution of eq.~(\ref{eq:Phieq}) can be found for ultrarelativistic
particles. In the limit $m\to0$, we can represent $\Phi=v\varphi$
in eq.~(\ref{eq:Phieq}), where $\varphi=\varphi(z)$ is a scalar
function and $v$ is a constant spinor satisfying $\Sigma_{3}v=\sigma v$
and $\gamma^{\bar{5}}v=\chi v$, with $\sigma=\pm1$ and $\chi=\pm1$,
since both $\Sigma_{3}$ and $\gamma^{\bar{5}}$ now commute with
the operator of eq.~(\ref{eq:Phieq}).%

We shall be interested in the description of left active neutrinos
satisfying $(1+\gamma^{\bar{5}})\psi$=0 and having $\chi=+1$. One
can show that, for right sterile neutrinos with $\chi=-1$, the interaction
with matter is washed out. Using the new variable $\rho=|V|az{}^{2}$
we can write the equation for $\varphi_{\sigma}$ as
%
\begin{equation}\label{eq:phisigma}
  \left[
    \rho\partial_{\rho}^{2}+\partial_{\rho}-\frac{\mu^2}{\rho} +
    \frac{\rho}{4}-\kappa
  \right]
  \varphi_{\sigma} = 0,
\end{equation}
where
\begin{equation}\label{eq:kappa}
  \kappa=\kappa_{0}+s\kappa_{1}-\mathrm{i}\frac{\sigma s}{4},
  \quad
  \kappa_{0} = \frac{p_{\perp}^{2}+m^{2}}{4|V|a},
  \quad
  \kappa_{1}=\frac{E}{2a},
  \quad
  \mu=\frac{1}{4}-\mathrm{i}\sigma\kappa_{1},
\end{equation}
and $s=\text{sgn}(V)=-1$ for neutrinos in a neutron matter; cf. eq.~(\ref{eq:Vnue}).%

\section{Classification of the neutrino quantum states\label{sec:CLASSQUANTST}}

In this section we shall find the solution of the wave equation for
neutrinos interacting with accelerated matter in the explicit form
and classify the neutrino quantum states.

The solutions of eq.~(\ref{eq:phisigma}) have the form,
\begin{equation}\label{eq:solMW}
  _{\zeta}\varphi_{\sigma} =
  \frac{1}{\sqrt{\rho}}M_{\mathrm{i}\zeta\kappa,\mu}(\zeta\mathrm{i}\rho),  
  \quad
  {}^{\zeta}\varphi_{\sigma} =
  \frac{1}{\sqrt{\rho}}W_{\mathrm{i}\zeta\kappa,\mu}(\mathrm{i}\zeta\rho),
  \quad
  \zeta=\pm,
\end{equation}
where $M_{\lambda,\mu}(z)$ and $W_{\lambda,\mu}(z)$ are the Whittaker
functions.
Note that solutions of the Dirac equation, expressed via Whittaker
functions, in the $1+1$ de Sitter space-time were also obtained in
ref.~\cite{Vil95}.

We shall suppose that the matter acceleration takes place in the spatial
interval $z_{\mathrm{in}}<z<z_{\mathrm{out}}$, where $z_{\mathrm{in}}\ll z_{\mathrm{out}}$.
It corresponds to the change of the dimensionless variable $\rho_{\mathrm{in}}<\rho<\rho_{\mathrm{out}}$,
where $\rho_{\mathrm{in}}\ll1$ and $\rho_{\mathrm{out}}\gg1$. Using
eq.~(\ref{eq:Mas0}), we get that $_{\zeta}\varphi$
can be identified with the quantum state corresponding to a definite
sign of the momentum projection along the $z$-axis at $\rho\sim\rho_{\mathrm{in}}$,
i.e. an in-state, whereas $^{\zeta}\varphi$ is the wave function
of this quantum state at $\rho\sim\rho_{\mathrm{out}}$. We shall
call it an out-state.

Therefore we can define two sets of the orthogonal neutrino wave functions: $_{\pm} \psi_n$ and $^{\pm} \psi_n$ (in and out solutions). These wave functions can be obtained from $^{\pm}\varphi$ and $_{\pm}\varphi$ using eq.~\eqref{eq:psipsir} and the relation between $\psi_z$ and $\Phi$ in section~\ref{sec:DIRACEQ}. We shall denote all the quantum numbers, like $p_{x,y}$, $E$, and $\sigma$, as $n$ for brevity.

The secondly quantized neutrino field $\hat{\psi}$ can be decomposed using $_{\pm} \psi_n$ and $^{\pm} \psi_n$ in the form,
\begin{equation}\label{eq:operdec}
  \hat{\psi} = \sum_n
  \left[
    \hat{a}_n(\text{in}) _{+} \psi_n + \hat{b}_n^\dagger(\text{in}) _{-} \psi_n
  \right] =
  \sum_n
  \left[
    \hat{a}_n(\text{out}) ^{+} \psi_n + \hat{b}_n^\dagger(\text{out}) ^{-} \psi_n
  \right],
\end{equation}
where $\hat{a}_n$ and $\hat{b}_n$ are the annihilation operators of the quantum state corresponding to a definite
sign of the momentum projection along the $z$-axis. These operators act on quantum states in ``in'' and ``out'' Fock spaces. The vacuum in these Fock spaces is defined as
\begin{equation}\label{eq:vac}
  \hat{a}_n(\text{in},\text{out})
  | 0 \rangle_\mathrm{in,out} =
  \hat{b}_n(\text{in},\text{out})
  | 0 \rangle_\mathrm{in,out} = 0.
\end{equation}
The creation and annihilation operators obey the usual anticommutation relations,
\begin{equation}\label{eq:anticom}
  \left[
    \hat{a}_n(\text{in},\text{out}),\hat{a}^\dagger_{n'}(\text{in},\text{out})
  \right]_+ =
  \left[
    \hat{b}_n(\text{in},\text{out}),\hat{b}^\dagger_{n'}(\text{in},\text{out})
  \right]_+ = \delta_{n,n'},
\end{equation}
with all the other anticommutators being equal to zero.

Now, let us find $_{\pm} \psi_n$ and $^{\pm} \psi_n$ in the explicit form. For this purpose we represent the bispinor $\psi_{z}$ of active ultrarelativistic neutrinos as $\psi_{z}^{\mathrm{T}}=(0,\eta)$. Here we assume the Dirac
matrices in the chiral representation~\cite{ItzZub80}
\begin{equation}\label{eq:chirrep}
  \gamma^{\bar{0}} =
  \left(
    \begin{array}{cc}
      0 & -1\\
      -1 & 0
    \end{array}
  \right),
  \quad
  \gamma^{\bar{k}} =
  \left(
    \begin{array}{cc}
      0 & \sigma_{k}\\
      -\sigma_{k} & 0
    \end{array}
  \right),
  \quad
  \gamma^{\bar{5}} =
  \left(
    \begin{array}{cc}
      1 & 0\\
      0 & -1
    \end{array}
  \right),
\end{equation}
where $\sigma_{k}=(\sigma_{1},\sigma_{2},\sigma_{3})$ are the Pauli
matrices. The two component spinors $\eta_{\sigma}$ corresponding
to the opposite helicity states with $\sigma=\pm1$, can be obtained
if we choose the following constant bispinors $v_{\sigma}$:
\begin{equation}\label{eq:vsigma}
  v_{+}^{\mathrm{T}}=(1,0,0,0),
  \quad
  v_{-}^{\mathrm{T}}=(0,1,0,0).
\end{equation}
Finally, using eqs.~(\ref{eq:chirrep}) and~(\ref{eq:vsigma}),
we get $\eta_{\pm}$ in the form,
\begin{align}\label{eq:etapmgen}
  \eta_{+} = & \Pi
  \left(
    \begin{array}{c}
      \varphi_{+} \\
      0
    \end{array}
  \right),
  \quad
  \eta_{-} = \Pi
  \left(
    \begin{array}{c}
      0 \\
      \varphi_{-}
    \end{array}
  \right),
\end{align}
where
\begin{align}\label{eq:projoper}
  \Pi = & 2\sqrt{a|V|}
  \left[
    s\frac{\sqrt{\rho}}{2}-\frac{E}{2a\sqrt{\rho}}+\mathrm{i}\sigma
    \left(
      \sqrt{\rho}\partial_{\rho}+\frac{1}{4\sqrt{\rho}}
    \right)
  \right] -
  \left(
    \bm{\sigma}_{\perp}\mathbf{p}_{\perp}
  \right),
\end{align}
is the projection operator, $\bm{\sigma}_{\perp}=\left(\sigma_{1},\sigma_{2}\right)$,
and $\mathbf{p}_{\perp}=\left(p_{x},p_{y}\right)$.

Let us first obtain out-states in the explicit form. Using eqs.~(\ref{eq:solMW}),
(\ref{eq:etapmgen}), and~(\ref{eq:projoper}), as well as eq.~(\ref{eq:WMder}),
we can obtain the wave functions corresponding to the negative momentum along the $z$-axis,
\begin{align}\label{eq:etaoutm}
  ^{-}\eta_{+} = & \frac{1}{\sqrt{\rho}}
  \left(
    \begin{array}{c}
      ^{-}\alpha_{+}
      W_{-1/4+\mathrm{i}\kappa_{0} - \mathrm{i}\kappa_{1},
      1/4-i\kappa_{1}}
      (i\rho) +
      {}^{-}\beta_{+}
      W_{-3/4+\mathrm{i}\kappa_{0}-\mathrm{i}\kappa_{1},
      -1/4-\mathrm{i}\kappa_{1}}
      (\mathrm{i}\rho) \\
      -\left(
        p_{x}+\mathrm{i}p_{y}
      \right)
      W_{-1/4+\mathrm{i}\kappa_{0}-\mathrm{i}\kappa_{1},
      1/4-\mathrm{i}\kappa_{1}}
      (\mathrm{i}\rho)
    \end{array}
  \right),
  \nonumber
  \\
  ^{-}\eta_{-} = & \frac{1}{\sqrt{\rho}}
  \left(
    \begin{array}{c}
      -\left(
        p_{x}-\mathrm{i}p_{y}
      \right)
      W_{1/4+\mathrm{i}\kappa_{0}-\mathrm{i}\kappa_{1},
      1/4+\mathrm{i}\kappa_{1}}
      (\mathrm{i}\rho)
      \\
      ^{-}\alpha_{-}
      W_{1/4+\mathrm{i}\kappa_{0}-\mathrm{i}\kappa_{1},
      1/4+\mathrm{i}\kappa_{1}} 
      (\mathrm{i}\rho) +
      {}^{-}\beta_{-}
      W_{3/4+i\kappa_{0}-\mathrm{i}\kappa_{1},
      -1/4+\mathrm{i}\kappa_{1}}
      (\mathrm{i}\rho)
    \end{array}
  \right).
\end{align}
Analogously one can derive the wave functions for the positive projection
of the momentum as
\begin{align}\label{eq:etaoutp}
  ^{+}\eta_{+}= & \frac{1}{\sqrt{\rho}}
  \left(
    \begin{array}{c}
      ^{+}\alpha_{+}
      W_{1/4-\mathrm{i}\kappa_{0}+\mathrm{i}\kappa_{1},
      1/4-\mathrm{i}\kappa_{1}}
      (-\mathrm{i}\rho) +
      {}^{+}\beta_{+}
      W_{3/4-\mathrm{i}\kappa_{0}+\mathrm{i}\kappa_{1},
      -1/4-\mathrm{i}\kappa_{1}}
      (-\mathrm{i}\rho)
      \\
      -\left(
        p_{x}+\mathrm{i}p_{y}
      \right)
      W_{1/4-\mathrm{i}\kappa_{0}+\mathrm{i}\kappa_{1},
      1/4-\mathrm{i}\kappa_{1}}
      (-\mathrm{i}\rho)
    \end{array}
  \right),
  \nonumber
  \\
  ^{+}\eta_{-} = & \frac{1}{\sqrt{\rho}}
  \left(
    \begin{array}{c}
      -\left(
        p_{x}-\mathrm{i}p_{y}
      \right)
      W_{-1/4-\mathrm{i}\kappa_{0}+\mathrm{i}\kappa_{1},
      1/4+\mathrm{i}\kappa_{1}}
      (-\mathrm{i}\rho)
      \\
      ^{+}\alpha_{-}
      W_{-1/4-\mathrm{i}\kappa_{0}+\mathrm{i}\kappa_{1},
      1/4+\mathrm{i}\kappa_{1}}
      (-\mathrm{i}\rho) +
      {}^{+}\beta_{-}
      W_{-3/4-\mathrm{i}\kappa_{0}+\mathrm{i}\kappa_{1},
      -1/4+\mathrm{i}\kappa_{1}}
      (-\mathrm{i}\rho)
    \end{array}
  \right).
\end{align}
The coefficients in eqs.~(\ref{eq:etainm}) and~(\ref{eq:etaoutp})
have the form,
\begin{align}
  ^{\zeta}\alpha_{\zeta'} = & -2E\sqrt{\frac{|V|}{a\rho}},  
  \quad
  {}^{\pm}\mathbf{\beta_{\pm}} = 2\sqrt{a|V|}e^{\mp3\pi\mathrm{i}/4},
  \notag
  \\
  {}^{\pm}\mathbf{\beta_{\mp}} = & 2\sqrt{a|V|}e^{\pm3\pi\mathrm{i}/4}
  \left(
    \kappa_{0}-2\kappa_{1}\mp\frac{\mathrm{i}}{2}
  \right),
\end{align}
where $\zeta$ and $\zeta'$ take the values $\pm$ independently.

We define the norm of the total wave function as
\begin{equation}\label{eq:scalprod}
  \left\langle
    \psi|\psi'
  \right\rangle =
  \int\mathrm{d}^{3}x\sqrt{-g}\bar{\psi}_{E,p_x,p_y}
  \gamma^{0}(x)\psi_{E',p_x',p_y'} =
  \left\Vert
    \psi
  \right\Vert^{2}
  \delta(E-E')\delta^{2}(\mathbf{p}_{\perp}-\mathbf{p}'_{\perp}),
\end{equation}
where $\bar{\psi} = \psi^\dagger \gamma^{\bar{0}}$. The norm of a two component spinor $\eta$ can be defined analogously. Using eqs.~(\ref{eq:etaoutm})
and~(\ref{eq:etaoutp}), we get that
\begin{equation}\label{eq:outconj}
  \sigma_{1}
  \left(
    ^{-}\eta_{-}
  \right)^{*} =
  {}^{+}\eta_{+},
  \quad
  \sigma_{1}
  \left(
    ^{+}\eta_{-}
  \right)^{*} =
  {}^{-}\eta_{+}.
\end{equation}
Therefore, one can establish the relation between the norms of the
spinors,
\begin{equation}\label{eq:outnorm}
  \left\Vert
    ^{-}\eta_{-}
  \right\Vert^{2} =
  \left\Vert
    ^{+}\eta_{+}
  \right\Vert^{2},
  \quad
  \left\Vert
    ^{+}\eta_{-}
  \right\Vert^{2} =
  \left\Vert
    ^{-}\eta_{+}
  \right\Vert^{2},
\end{equation}
Note that the analog of eqs.~(\ref{eq:outconj}) and~(\ref{eq:outnorm})
was found in ref.~\cite{HaoChe13}.

The in-states can be obtained in the same manner as eqs.~(\ref{eq:etaoutm})
and~(\ref{eq:etaoutp}). We just give the final result,
\begin{align}\label{eq:etainm}
  _{-}\eta_{+} = & \frac{1}{\sqrt{\rho}}
  \left(
    \begin{array}{c}
      _{-}\alpha_{+}
      M_{-1/4+\mathrm{i}\kappa_{0}-\mathrm{i}\kappa_{1},
      1/4-\mathrm{i}\kappa_{1}}
      (\mathrm{i}\rho) +
      _{-}\beta_{+}
      M_{-3/4+\mathrm{i}\kappa_{0}-\mathrm{i}\kappa_{1},
      -1/4-\mathrm{i}\kappa_{1}}
      (\mathrm{i}\rho)
      \\
      -\left(
        p_{x}+\mathrm{i}p_{y}
      \right)
      M_{-1/4+\mathrm{i}\kappa_{0}-\mathrm{i}\kappa_{1},
      1/4-\mathrm{i}\kappa_{1}}
      (\mathrm{i}\rho)
    \end{array}
  \right),
  \nonumber
  \\
  _{-}\eta_{-} = & \frac{1}{\sqrt{\rho}}
  \left(
    \begin{array}{c}
      -\left(
        p_{x}-\mathrm{i}p_{y}
      \right)
      M_{1/4+\mathrm{i}\kappa_{0}-\mathrm{i}\kappa_{1},
      1/4+\mathrm{i}\kappa_{1}}
      (\mathrm{i}\rho)
      \\
      _{-}\alpha_{-}
      M_{1/4+\mathrm{i}\kappa_{0}-\mathrm{i}\kappa_{1},
      1/4+\mathrm{i}\kappa_{1}}   
      (\mathrm{i}\rho) +
      {}_{-}\beta_{-}
      M_{3/4+\mathrm{i}\kappa_{0}-\mathrm{i}\kappa_{1},
      -1/4+\mathrm{i}\kappa_{1}}
      (\mathrm{i}\rho)
    \end{array}
  \right),
\end{align}
and
\begin{align}\label{eq:etainp}
  _{+}\eta_{+} = & \frac{1}{\sqrt{\rho}}
  \left(
    \begin{array}{c}
      _{+}\alpha_{+}
      M_{1/4-\mathrm{i}\kappa_{0}+\mathrm{i}\kappa_{1},
      1/4-\mathrm{i}\kappa_{1}}
      (-\mathrm{i}\rho) +
      {}_{+}\beta_{+}
      M_{3/4-\mathrm{i}\kappa_{0}+\mathrm{i}\kappa_{1},
      -1/4-\mathrm{i}\kappa_{1}}
      (-\mathrm{i}\rho) \\
      -\left(
        p_{x}+\mathrm{i}p_{y}
      \right)
      M_{1/4-\mathrm{i}\kappa_{0}+\mathrm{i}\kappa_{1},
      1/4-\mathrm{i}\kappa_{1}}
      (-\mathrm{i}\rho)
    \end{array}
  \right),
  \nonumber
  \\
  _{+}\eta_{-}= & \frac{1}{\sqrt{\rho}}
  \left(
    \begin{array}{c}
      -\left(
        p_{x}-\mathrm{i}p_{y}
      \right)
      M_{-1/4-\mathrm{i}\kappa_{0}+\mathrm{i}\kappa_{1},
      1/4+\mathrm{i}\kappa_{1}}
      (-\mathrm{i}\rho)
      \\
      _{+}\alpha_{-}
      M_{-1/4-\mathrm{i}\kappa_{0}+\mathrm{i}\kappa_{1},
      1/4+\mathrm{i}\kappa_{1}}    
      (-\mathrm{i}\rho) +
      {}_{+}\beta_{-}
      M_{-3/4-\mathrm{i}\kappa_{0}+\mathrm{i}\kappa_{1},
      -1/4+\mathrm{i}\kappa_{1}}
      (-\mathrm{i}\rho)
    \end{array}
  \right).
\end{align}
The coefficients in eqs.~(\ref{eq:etainm}) and~(\ref{eq:etainp}) are
\begin{align}
  _{\zeta}\alpha_{\zeta'} = & -2E\sqrt{\frac{|V|}{a\rho}},
  \quad
  {}_{-}\mathbf{\beta_{\pm}} = e^{\pi\mathrm{i}/4}\sqrt{a|V|}
  \left(
    4\kappa_{1}\pm\mathrm{i}
  \right),
  \notag
  \\
  {}_{+}\mathbf{\beta_{\pm}} = &
  e^{-\pi\mathrm{i}/4}\sqrt{a|V|}
  \left(
    4\kappa_{1}\pm\mathrm{i}
  \right),
\end{align}
where $\zeta$ and $\zeta'$ again take the values $\pm$ independently.
Analogously to the out-states, we can establish the following relations
between the norms of the wave functions: $\left\Vert _{-}\eta_{-}\right\Vert^2 = \left\Vert _{+}\eta_{+}\right\Vert^2$
and $\left\Vert _{+}\eta_{-}\right\Vert^2 =\left\Vert _{-}\eta_{+}\right\Vert^2$.
These expressions are valid since it follows from eqs.~(\ref{eq:etainm})
and~(\ref{eq:etainp}) that $\sigma_{1}\left(_{-}\eta_{-}\right)^{*}={}_{+}\eta_{+}$
and $\sigma_{1}\left(_{+}\eta_{-}\right)^{*}={}_{-}\eta_{+}$.

To complete the determination of the states we need to establish the
relation between the norms of the wave functions for the same helicity
and different signs of the momentum projection. This relation
can be found in the case when $\kappa_{1}\gg\kappa_{0}$. Using eqs.~(\ref{eq:mupm1/2})
and~(\ref{eq:Gammanorm}), we get the ratio of the norms of $^{+}\eta_{-}$
and $^{-}\eta_{-}$ as
%
\begin{equation}\label{eq:normrat}
  \frac{
  \left\Vert
    ^{+}\eta_{-}
  \right\Vert^{2}
  }
  {
  \left\Vert
    ^{-}\eta_{-}
  \right\Vert ^{2}
  } =
  \frac{\pi e^{-2\pi\kappa_{1}}}{\cosh2\pi\kappa_{1}}.
\end{equation}
Note that one can derive the analogous relation between, e.g., the wave
functions $_{+}\eta_{-}$ and $_{-}\eta_{-}$. However, we shall not
present it here.

\section{Particle creation in an accelerated matter\label{sec:PAIRSCR}}

In this section we describe the creation of $\nu\bar{\nu}$ pairs
in a background matter moving with a linear acceleration. We also reproduce
the Unruh effect.

Both out- and in-states are the complete sets of orthogonal wave functions.
Thus we can expand any in-state wave function over the system of out-states,
\begin{equation}\label{eq:inoutrelgen}
  _{\zeta}\eta_{\sigma} =
  \sum_{\zeta'=\pm}
  G
  \left(
    ^{\zeta'} | {}_{\zeta}
  \right)
  {}^{\zeta'}\eta_{\sigma},
  \quad
  \zeta=\pm,
\end{equation}
where the Bogoliubov coefficients $G\left(^{\zeta'}|{}_{\zeta}\right)$
satisfy the unitarity conditions,
\begin{equation}\label{eq:unitcond}
  \sum_{\zeta=\pm} G
  \left(
    ^{\zeta'}|{}_{\zeta}
  \right)
  G
  \left(
    _{\zeta}|^{\zeta''}
  \right) =
  \sum_{\zeta=\pm}
  G
  \left(
    _{\zeta'}|^{\zeta}
  \right)
  G
  \left(
    ^{\zeta}|{}_{\zeta''}
  \right) =
  \delta_{\zeta'\zeta''}.
\end{equation}
Here $G\left(_{\zeta}|^{\zeta'}\right)=G\left(^{\zeta'}|{}_{\zeta}\right)^{*}$
are the coefficients of the inverse transformation.

The relation between in and out wave finctions in eq.~\eqref{eq:inoutrelgen} is equivalent to the Bogoliubov transformation of the creation and annihilation operators defined in eqs.~\eqref{eq:operdec}-\eqref{eq:anticom},
\begin{align}
  \hat{a}_n(\mathrm{out}) = &
  G\left(^{+}|{}_{+}\right) \hat{a}_n(\mathrm{in}) +
  G\left(^{+}|{}_{-}\right) \hat{b}_n^\dagger(\mathrm{in}),
  \notag
  \\
  \hat{b}^\dagger_n(\mathrm{out}) = &
  G\left(^{-}|{}_{+}\right) \hat{a}_n(\mathrm{in}) +
  G\left(^{-}|{}_{-}\right) \hat{b}_n^\dagger(\mathrm{in}).
\end{align}
Therefore, the nonzero mean value of the number density operator of out-states over the ``in'' vacuum, $_\mathrm{in} \langle 0 | \hat{a}^\dagger_n(\text{out}) \hat{a}_n(\text{out}) | 0 \rangle_\mathrm{in} = \left| G\left( ^{+} | _{-} \right) \right|^2 \neq 0$, can be interpreted as the creation of a $\nu\bar{\nu}$ pair. 

From the formal point of view the vacuum instability leading to the $\nu\bar{\nu}$ pairs creation, described in the present work, is analogous to the pairs creation by an external field depending on spatial coordinates. It should be noted that a more detailed analysis of this problem was recently 
presented in refs.~\cite{GavGit13,GavGit15}.

Setting $\zeta=-$ in eq.~(\ref{eq:inoutrelgen}) we get that
\begin{align}\label{eq:etainmpart}
  _{-}\eta_{\sigma} = & G
  \left(
    ^{+}|{}_{-}
  \right)
  {}^{+}\eta_{\sigma} + G
  \left(
    ^{-}|{}_{-}
  \right)
  {}^{-}\eta_{\sigma}.
\end{align}
The Bogoliubov coefficients in eq.~(\ref{eq:etainmpart}) are related
to the scalar products, defined in eq.~(\ref{eq:scalprod}), of the
corresponding in- and out-states as
\begin{equation}\label{eq:Gpmmm}
  G
  \left(
    ^{+}|{}_{-}
  \right) =
  \left\langle
    ^{+}\eta_{\sigma}|{}_{-}\eta_{\sigma}
  \right\rangle,
  \quad
  G
  \left(
    ^{-}|{}_{-}
  \right) =
  \left\langle
    ^{-}\eta_{\sigma}|{}_{-}\eta_{\sigma}
  \right\rangle.
\end{equation}
Since we study ultrarelativistic particles, there is a strong correlation
between their momentum and spin: neutrinos are left-polarized, whereas antineutrinos
are right-polarized. If $\sigma=-1$, the nonzero coefficient $G\left(^{+}|{}_{-}\right)$
corresponds to the creation of the pair of $\nu$ with $(p_{z}>0,\sigma=-1)$
and $\bar{\nu}$ with $(p_{z}<0,\sigma=-1)$. In case $\sigma=+1$,
$G\left(^{+}|{}_{-}\right)$ describes the creation of $\nu$ with
$(p_{z}<0,\sigma=+1)$ and $\bar{\nu}$ with $(p_{z}>0,\sigma=+1)$.
However, these two sets of states are identical. Thus, if we define
the corresponding probabilities
\begin{equation}\label{eq:probdef}
  P_{\sigma} =
  \left|
    G
    \left(
      ^{+}|{}_{-}
    \right)
  \right|^{2},
\end{equation}
we should get that $P_{-}+P_{+}=1$. In fact, this relation between
the probabilities is the consequence of eq.~(\ref{eq:outconj}).

To derive $P_{\sigma}$ in the explicit form, we should use eq.~(\ref{eq:MWrel}).
For $\sigma=-1$, on the basis of eqs.~\eqref{eq:etaoutm}, \eqref{eq:etaoutp}, and~\eqref{eq:etainm}, we have
\begin{align}\label{eq:etaoutmm}
  _{-}\eta_{-} = & \Gamma
  \left(
    \frac{3}{2}+2\mathrm{i}\kappa_{1}
  \right)
  \exp
  \left[
    \pi
    \left(
      \kappa_{0}-\kappa_{1}
    \right) -
    \frac{\mathrm{i}\pi}{4}
  \right]
  \notag
  \\
  & \times
  \left[
    \frac{\exp
    \left(
      -\pi\kappa_{1} + 3\pi \mathrm{i}/4
    \right)
    }{
    \Gamma
    \left(
      1+\mathrm{i}\kappa_{0}
    \right)
    }
    {}^{-}\eta_{-}+\frac{1}{\Gamma
    \left(
      1/2+2\mathrm{i}\kappa_{1}-\mathrm{i}\kappa_{0}
    \right)
    }
    {}^{+}\eta_{-}
  \right].
\end{align}
Substituting eq.~(\ref{eq:etaoutmm}) to eq.~(\ref{eq:Gpmmm}) we
obtain
\begin{equation}\label{eq:ratm}
  \frac{G
  \left(
    ^{-}|{}_{-}
  \right)
  }{
  G
  \left(
    ^{+}|{}_{-}
  \right)
  } = \exp
  \left(
    -\pi\kappa_{1}+\frac{3\pi \mathrm{i}}{4}
  \right)
  \frac{
  \left\Vert
    ^{-}\eta_{-}
  \right\Vert^{2}
  }{
  \left\Vert
    ^{+}\eta_{-}
  \right\Vert^{2}
  }
  \frac{\Gamma
  \left(
    1/2+2\mathrm{i}\kappa_{1}-i\kappa_{0}
  \right)
  }{
  \Gamma
  \left(
    1+\mathrm{i}\kappa_{0}
  \right)
  }.
\end{equation}
Analogously, if $\sigma=+1$, one gets that
\begin{align}\label{eq:etaoutmp}
  _{-}\eta_{+}  = & \Gamma
  \left(
    \frac{3}{2}-2\mathrm{i}\kappa_{1}
  \right)
  \exp
  \left[
    \pi
    \left(
      \kappa_{0}-\kappa_{1}
    \right)+\frac{\mathrm{i}\pi}{4}
  \right]
  \notag
  \\
  & \times
  \left[
    \frac{\exp
    \left(
      \pi\kappa_{1}+3\pi \mathrm{i}/4
    \right)
    }{
    \Gamma
    \left(
      1/2-2\mathrm{i}\kappa_{1}+\mathrm{i}\kappa_{0}
    \right)}
    {}^{-}\eta_{+}+\frac{1}{\Gamma
    \left(
      1-\mathrm{i}\kappa_{0}
    \right)}{}^{+}\eta_{+}
  \right],
\end{align}
and
\begin{equation}\label{eq:ratp}
  \frac{G
  \left(
    ^{-}|{}_{-}
  \right)
  }{
  G
  \left(
    ^{+}|{}_{-}
  \right)} =
  \exp
  \left(
    \pi\kappa_{1} + \frac{3\pi \mathrm{i}}{4}
  \right)
  \frac{
  \left\Vert
    ^{-}\eta_{+}
  \right\Vert^{2}
  }{
  \left\Vert
    ^{+}\eta_{+}
  \right\Vert^{2}}
  \frac{\Gamma
    \left(
      1-\mathrm{i}\kappa_{0}
    \right)
    }{
    \Gamma
    \left(
      1/2-2\mathrm{i}\kappa_{1}+\mathrm{i}\kappa_{0}
    \right)}.
\end{equation}
%
%
Finally, accounting for eqs.~(\ref{eq:unitcond}), (\ref{eq:probdef}),
and~(\ref{eq:Gammanorm}), we get that
\begin{align}\label{eq:Pmp}
  P_{-}= &
  \left[
    1+e^{-2\pi\kappa_{1}}
    \frac{
    \left\Vert
      ^{-}\eta_{-}
    \right\Vert^{4}}
    {
    \left\Vert
      ^{+}\eta_{-}
    \right\Vert^{4}
    }
    \frac{\sinh\pi\kappa_{0}}
    {\kappa_{0}\cosh\pi(2\kappa_{1}-\kappa_{0})}
  \right]^{-1},
  \nonumber
  \\
  P_{+}= &
  \left[
    1+e^{2\pi\kappa_{1}}
    \frac{
    \left\Vert
      ^{-}\eta_{+}
    \right\Vert^{4}}
    {
    \left\Vert
      ^{+}\eta_{+}
    \right\Vert^{4}}
    \frac{\kappa_{0}\cosh\pi(2\kappa_{1}-\kappa_{0})}
    {\sinh\pi\kappa_{0}}
  \right]^{-1}.
\end{align}
Taking into account eq.~\eqref{eq:outnorm}, one can see that $P_{-}+P_{+}=1$, as it should be.

Let us discuss the situation when $\kappa_{1}\gg1\gg\kappa_{0}$,
i.e. the creation of $\nu\bar{\nu}$ pairs is mainly owing to the
noninertial effects. We can define the number of created pairs per
second in the energy interval $\mathrm{d}E$ as $\mathrm{d}\dot{N} = P_{-} \mathrm{d}E/2\pi$.
On the basis of eqs.~\eqref{eq:normrat} and~(\ref{eq:Pmp}), we get the distribution of created
particles in this limit as
\begin{equation}\label{eq:endistr}
  \frac{\mathrm{d}\dot{N}}{\mathrm{d}E} =
  \exp
  \left(
    -\frac{E+\delta E}{T_{\mathrm{eff}}}
  \right),
  \quad
  T_{\mathrm{eff}} = \frac{a}{2\pi},
  \quad
  \delta E = \frac{p_{\perp}^{2}+m^{2}}{8|V|}.
\end{equation}
Here we reproduce the temperature of the Unruh radiation $T_{\mathrm{eff}}$~\cite{Unr76}. The correction to the Unruh effect owing to the specific electroweak neutrino radiation is given by the quantity $\delta E$ in eq.~(\ref{eq:endistr}). Note that eq.~(\ref{eq:endistr})
is obtained under the assumption that $E\gg\delta E$.

%

\section{Neutrino pairs creation in SN\label{sec:APPL}}

In this section we shall discuss a possible application of the obtained
results for the description of the $\nu\bar{\nu}$ pairs creation
in a core collapsing SN. In particular, we shall obtain the
astrophysical upper limit on the neutrino mass.

The distribution of created $\nu\bar{\nu}$ pairs is given in eq.~(\ref{eq:endistr}).
If we discuss the case of the neutrinos propagation along the $z$-axis,
the correction to the Unruh effect does not suppress the pair creation
if $\delta E\ll T_{\mathrm{eff}}$. This condition can be rewritten
as the constraint on the neutrino mass: $m\ll m_{\mathrm{cr}}$, where
\begin{equation}\label{eq:masslim}
  m_{\mathrm{cr}}=2\sqrt{\frac{|V|a}{\pi}}.
\end{equation}
Note that the constraint in eq.~(\ref{eq:masslim}) is theoretical.
It means that, if $\nu\bar{\nu}$ pairs, created because of the vacuum instability in an accelerated matter,
are observed experimentally, then the neutrino mass should be less than $m_{\mathrm{cr}}$
in eq.~(\ref{eq:masslim}).

Let us evaluate $m_{\mathrm{cr}}$ for a core collapsing SN. When
the plasma pressure cannot equilibrate the gravity of the upper layers
of a protostar, the radius of a protostar core starts to decrease
until the matter density reaches $\rho_{n}\sim10^{14}\thinspace\text{g}\cdot\text{cm}^{-3}$.
Assuming that PNS mainly consists of neutrons, this value corresponds
to the neutron density $n_{n}\approx6\times10^{37}\thinspace\text{cm}^{-3}$.
At this point no further core compression happens and initially falling
matter bounces from a dense core which then turns out to be PNS.
According to the results of ref.~\cite{ThoBurPin03}, the matter velocity changes by $|\Delta v|\approx5\times10^{9}\thinspace\text{cm}\cdot\text{s}^{-1}$
during $\Delta t\approx1\thinspace\text{ms}$ after a bounce. It
gives the nonzero matter acceleration $a=|\Delta v|/\Delta t\approx5\times10^{12}\thinspace\text{cm}\cdot\text{s}^{-2}$.
Using eq.~(\ref{eq:Vnue}) and the chosen values of all the parameters,
we get that $m_{\mathrm{cr}}\approx7.2\times10^{-7}\thinspace\text{eV}$.
The obtained value of $m_{\mathrm{cr}}$ is comparable with the (theoretical)
upper bounds on the neutrino masses derived in ref.~\cite{DvoGavGit14}.

The total number of $\nu\bar{\nu}$ pairs emitted during $\Delta t$ from the whole surface of PNS can be obtained by integrating eq.~\eqref{eq:endistr} over $E$ and $\mathbf{p}_\perp$. The final result reads
\begin{equation}\label{eq:numpart}
  N = 8 |V| T_\mathrm{eff}^2 \Delta t R^2
  \exp
  \left(
    -\frac{m^{2}}{8|V|T_\mathrm{eff}}
  \right),
\end{equation}
where $R$ is the PNS radius. The maximal matter acceleration at a bounce happens at $R \sim 10^3\thinspace\text{km}$~\cite{ThoBurPin03}. Therefore, assuming that $m \sim m_\mathrm{cr}$, on the basis of eq.~\eqref{eq:numpart} one gets that $N \sim 10^{11}$ neutrino pairs can be emitted by PNS.

This $\nu\bar{\nu}$ flux is much less than the flux of $\nu_e$ emitted in a neutrino burst, which can be estimated as $N_{\nu_e} \sim 10^{54}$~\cite{GiuKim07p517}. However, the typical energy of $\nu$ and $\bar{\nu}$, created owing to the vacuum instability in a dense nuclear matter in PNS, is in the eV range; cf. ref.~\cite{DvoGavGit14}, and $E_{\nu_e} \sim 10\thinspace\text{MeV}$. Therefore, at the bounce stage, high energy $\nu_e$'s will be absorbed by the dense matter of PNS whereas low energy  $\nu$ and $\bar{\nu}$ will freely escape PNS. Indeed, using the neutrino scattering cross sections given in ref.~\cite{GiuKim07p160}, one finds that the mean free path of $\nu$ and $\bar{\nu}$, with energies in eV range, in background matter with the density $\rho_{n}\sim10^{14}\thinspace\text{g}\cdot\text{cm}^{-3}$ is about $10^{10}\thinspace\text{km}$, which is far beyond $R = 10^3\thinspace\text{km}$. Thus one can consider the emission of $\nu\bar{\nu}$ pairs created by the accelerated matter in PNS as a precursor of the neutrino burst.

Nevertheless, it is unlikely that such $\nu$ and $\bar{\nu}$, created by the proposed mechanism, can be detected using modern experimental techniques. Despite the recent suggestions to observe nonrelativistic relic neutrinos with sub-eV energies~\cite{LonLunSab14}, the flux of $\nu\bar{\nu}$ pairs emitted in a SN explosion is quite low to be detected in any existing neutrino telescope even if this explosion happens in our Galaxy.


\section{Conclusion\label{sec:CONCL}}

In conclusion we mention that, in the present work, we have studied
the evolution of neutrinos electroweakly interacting with a linearly
accelerated background matter. We have briefly reminded the standard
model neutrino interaction with background fermions in the flat space-time
in section~\ref{sec:NUMATTFLAT}. Then, in section~\ref{sec:DIRACEQ},
we have derived the Dirac equation for a massive neutrino, electroweakly
interacting with background fermions, in the Rindler space-time corresponding
to a linearly accelerated matter. The solution of this Dirac equation
has been found for ultrarelativistic particles. The quantum states
of neutrinos have been analyzed in section~\ref{sec:CLASSQUANTST}.
In section~\ref{sec:PAIRSCR}, we have applied our results for the
description of the $\nu\bar{\nu}$ pairs creation in an accelerated
matter. Finally, in section~\ref{sec:APPL}, we have considered the
$\nu\bar{\nu}$ pairs creation in a collapsing SN at the bounce stage.

As a main tool in our study we have used the method of exact solutions
of the Dirac equation for a massive neutrino rewritten in a comoving
frame, where matter is at rest. Firstly, this method enables one to
unambiguously determine the neutrino interaction potentials with background
fermions which move with an acceleration~\cite{Dvo14,Dvo15}. Secondly,
unlike refs.~\cite{FraNau13,StuTok14}, where the matter acceleration
was accounted for only in the neutrino refraction index, our approach
allowed us to examine noninertial effects.

We have demonstrated that the neutrino interaction with a linearly
accelerated matter can induce the vacuum instability which leads to
the creation of $\nu\bar{\nu}$ pairs. We have discussed the situation when an observer was
in the accelerated frame since we used the coordinates in the Rindler wedge. Therefore, the main contribution to the
$\nu\bar{\nu}$ pairs creation is owing to the Unruh effect~\cite{Unr76}.
We have rederived the temperature of the Unruh thermal radiation and
studied the correction to the Unruh effect due to the specific electroweak
neutrino interaction.

The electroweak decays of accelerated particles were studied refs.~\cite{VanMat01,SuzYam03}. The probability of the decay of an accelerated proton $p \xrightarrow{a} e^+ n \nu_e$ was shown in refs.~\cite{VanMat01,SuzYam03} to be nonzero. In those works the hadronic current was described quasiclassically in the flat space-time. Moreover it was demonstrated that the decay rate is equal to the sum of the decay rates of cross symmetric processes. These cross symmetric processes were studied using the exact solutions of the Dirac equation in the Rindler wedge. The results of refs.~\cite{VanMat01,SuzYam03} may be interpreted as the indirect proof of the existence of the Unruh effect.

Unlike refs.~\cite{VanMat01,SuzYam03}, in the present work we consider the elastic forward neutrino scattering off accelerated background fermions $f$ with a possible creation of $\nu \bar{\nu}$ pairs: $\nu f \xrightarrow{a} \nu f (\nu\bar{\nu})$. We treat background fermions as a classical external current. Therefore we confirm the result of refs.~\cite{VanMat01,SuzYam03} that the Unruh effect for neutrinos takes place if background fermions are treated quasiclassically.

In our analysis we use the exact solution of the Dirac equation which accounts for both the interaction with matter and its acceleration, i.e. our results exactly take into account all terms in the perturbative expansion over $G_\mathrm{F}$. It is the main difference from refs.~\cite{VanMat01,SuzYam03}, where the electroweak decays were studied on the tree level linear in $G_\mathrm{F}$. The chosen way to describe neutrinos interacting with an accelerated matter allowed us not only establish the existence of the Unruh effect for these particles but also to obtain the correction to this effect in eq.~\eqref{eq:endistr}.

The obtained electroweak correction to the Unruh effect was shown
to contribute to the $\nu\bar{\nu}$ pairs creation in a core
collapsing SN. Assuming that the process of the pairs creation is
not suppressed, we have obtained the upper (theoretical) limit to
the neutrino mass; cf. eq.~(\ref{eq:masslim}). This constraint is
in agreement with the result of ref.~\cite{DvoGavGit14}. Unfortunately, nowadays it is almost impossible to experimentally detect $\nu$ and $\bar{\nu}$ created in the proposed mechanism. 

It should be noted that previously the creation of $\nu\bar{\nu}$ pairs in dense matter of a neutron star (NS) was considered in refs.~\cite{Kah98,KusPos02,Koe05} using the Schwinger mechanism. For the first time the study of this problem based of the quantum field theory was undertaken in ref.~\cite{DvoGavGit14}. However, in ref.~\cite{DvoGavGit14} the case of the time dependent effective potential of the neutrino interaction with the background matter was discussed. Thus the model considered in ref.~\cite{DvoGavGit14} corresponds to the neutronization stage of PNS.

The process of the $\nu\bar{\nu}$ pairs creation described in the present work can be attributed to the dependence of the external field (background matter) on spatial coordinates. The general description of the vacuum instability in case of coordinate dependent external fields was recently discussed in refs.~\cite{GavGit13,GavGit15}. In our work for the first time we consider the  coordinate dependent effective potential of the neutrino interaction with background matter as a source of the neutrino vacuum instability.

Finally we mention that the creation of $\nu\bar{\nu}$ pairs considered in the present paper is different from the bremsstrahlung emission of $\nu\bar{\nu}$ pairs in the NS matter; cf. refs.~\cite{GiuKim07p517,Raf96}. Neutrino pairs emitted in the latter effect have energies proportional to the temperature of the nuclear matter in NS and thus correspond to high energies. On the contrary, we predict the coherent emission of low energy particles.

\acknowledgments{
I am thankful to V.G.~Bagrov, S.P.~Gavrilov, and S.~Haouat for useful discussions, to A.I.~Studenikin
for communications,
to the Tomsk State University Competitiveness Improvement Program and to RFBR (research project No.~15-02-00293) for partial support.}

\appendix

\section{Properties of Whittaker functions\label{sec:WMPROP}}

In this appendix we list some of the useful properties of Whittaker
and Euler gamma functions. These expressions are taken from ref.~\cite{GraRuz07}.

The asymptotic expansions of Whittaker functions read
\begin{alignat}{2}\label{eq:Mas0}
  M_{\lambda,\mu}(z) & \sim z^{1/2+\mu}+\dotsc, &
  \quad
  & \text{at}
  \quad  
  |z| \to 0,
  \\
  W_{\lambda,\mu}(z) & \sim z^{\lambda}e^{-z/2}+\dotsc, &
  \quad
  & \text{at}
  \quad  
  |z| \to \infty.
\end{alignat}
The derivatives of the Whittaker functions have the form,
\begin{align}\label{eq:WMder}
  \frac{\mathrm{d}W_{\pm i\kappa,\mu}(\pm\mathrm{i}\rho)}
  {\mathrm{d}\rho} = &
  \frac{e^{\pm3\mathrm{i}\pi/4}}{\sqrt{\rho}}
  \left[
    \kappa\mp\mathrm{i}
    \left(
      \mu-\frac{1}{2}
    \right)
  \right]
  W_{\pm\mathrm{i}\kappa-1/2,\mu-1/2}(\pm\mathrm{i}\rho)
  \nonumber
  \\
  & -
  \left(
    \frac{2\mu-1}{2\rho}\pm\frac{\mathrm{i}}{2}
  \right)
  W_{\pm\mathrm{i}\kappa,\mu}(\pm\mathrm{i}\rho),
  \nonumber
  \\
  \frac{\mathrm{d}M_{\pm\mathrm{i}\kappa,\mu}(\pm\mathrm{i}\rho)}
  {\mathrm{d}\rho} = &
  \frac{2\mu e^{\pm\mathrm{i}\pi/4}}{\sqrt{\rho}}
  M_{\pm\mathrm{i}\kappa-1/2,\mu-1/2}(\pm\mathrm{i}\rho) -
  \left(
    \frac{2\mu-1}{2\rho}\pm\frac{\mathrm{i}}{2}
  \right)
  M_{\pm\mathrm{i}\kappa,\mu}(\pm\mathrm{i}\rho),
  \nonumber
  \\
  \frac{\mathrm{d}W_{\pm\mathrm{i}\kappa,\mu}(\pm\mathrm{i}\rho)}
  {\mathrm{d}\rho} = &
  -\frac{e^{\pm\mathrm{i}\pi/4}}{\sqrt{\rho}}
  W_{\pm\mathrm{i}\kappa+1/2,\mu-1/2}(\pm\mathrm{i}\rho) -   
  \left(
    \frac{2\mu-1}{2\rho}\mp\frac{\mathrm{i}}{2}
  \right)
  W_{\pm\mathrm{i}\kappa,\mu}(\pm\mathrm{i}\rho),
  \nonumber
  \\
  \frac{\mathrm{d}M_{\pm\mathrm{i}\kappa,\mu}(\pm\mathrm{i}\rho)}
  {\mathrm{d}\rho} = &
  \frac{2\mu e^{\pm\mathrm{i}\pi/4}}{\sqrt{\rho}}
  M_{\pm\mathrm{i}\kappa+1/2,\mu-1/2}(\pm\mathrm{i}\rho) -
  \left(
    \frac{2\mu-1}{2\rho}\mp\frac{\mathrm{i}}{2}
  \right)
  M_{\pm\mathrm{i}\kappa,\mu}(\pm\mathrm{i}\rho).
\end{align}
%
At $\mu=\lambda\pm1/2$, $W_{\lambda,\mu}(z)$ takes the form,
\begin{equation}\label{eq:mupm1/2}
  W_{\lambda,\lambda-1/2}(z) = z^{\lambda}e^{-z/2},
  \quad
  W_{\lambda,\lambda+1/2}(z) = z^{-\lambda}e^{z/2}
  \Gamma(1+2\lambda,z),
\end{equation}
where $\Gamma(a,z)$ is the incomplete gamma function.

The function $M_{\lambda,\mu}(z)$ is related to $W_{\lambda,\mu}(z)$
by the following expression:
\begin{equation}\label{eq:MWrel}
  M_{\lambda,\mu}(z) =
  \Gamma
  \left(
    2\mu+1
  \right)
  e^{-\mathrm{i}\pi\lambda}
  \left[
    \frac{W_{-\lambda,\mu}
    \left(
      e^{-\mathrm{i}\pi}z
    \right)
    }{
    \Gamma
    \left(
      1/2+\mu-\lambda
    \right)
    } +
    e^{\mathrm{i}\pi(\mu+1/2)}\frac{W_{\lambda,\mu}
    \left(
      z
    \right)
    }{
    \Gamma
    \left(
      1/2+\mu+\lambda
    \right)
    }
  \right],
\end{equation}
which is valid if $-\pi/2<\arg z<3\pi/2$ and $2\mu\neq-1,-2,\dotsc.$

The absolute value of the Euler gamma function of the complex
argument in some cases can be evaluated as
\begin{equation}\label{eq:Gammanorm}
  \left|
    \Gamma
    \left(
      \frac{1}{2}+\mathrm{i}x
    \right)
  \right|^{2} =
  \frac{\pi}{\cosh\pi x},
  \quad
  \left|
    \Gamma
    \left(
      1+\mathrm{i}x
    \right)
  \right|^{2} =
  \frac{\pi x}{\sinh\pi x},
\end{equation}
where $x$ is a real parameter.

\end{document}